\documentclass[12pt]{article}
\usepackage[latin1]{inputenc}
\usepackage{amsmath,amssymb}
\usepackage{a4}
\usepackage{latexsym,cite}
\newcommand{\dete}{\boldsymbol{e}}
\begin{document}
\begin{titlepage}
\title{\begin{flushright}
  {\normalsize TUW-04-35\\
    LU-ITP 2004/045\\[-2ex]
  hep-th/0412007}
\end{flushright}\vspace{1cm}
Classical and Quantum Integrability of 2D Dilaton
Gravities in Euclidean space
}
\author{L.~Bergamin\thanks{{\tt bergamin@tph.tuwien.ac.at}}~,
  D.~Grumiller\thanks{{\tt grumiller@itp.uni-leipzig.de}}~,
  W.~Kummer\thanks{{\tt wkummer@tph.tuwien.ac.at}}~ and
  D.V.~Vassilevich\thanks{{\tt Dmitri.Vassilevich@itp.uni-leipzig.de}}}
\date{February 3, 2005}
\maketitle
\renewcommand{\thefootnote}{\fnsymbol{footnote}}
\begin{center}

{\footnotemark[1]\footnotemark[2]\footnotemark[3]Institute for Theoretical Physics, TU Vienna,\\ Wiedner Hauptstr.~8--10/136, A-1040 Vienna, Austria}

\vspace{0.5cm}

{\footnotemark[2]\footnotemark[4]Institute for Theoretical Physics, University of Leipzig\\Augustusplatz 10--11, D-04103 Leipzig, Germany}

\vspace{0.5cm}

{\footnotemark[4]V.A.\ Fock Institute of Physics, St.~Petersburg University, Russia}

\vspace{0.5cm}

\end{center}
\abstract{Euclidean dilaton gravity in two dimensions is studied exploiting 
its representation as a complexified first order gravity model. 
All local classical solutions are obtained. A global discussion reveals that for a 
given model only a restricted class of topologies is consistent with the 
metric and the dilaton. A particular case of string motivated Liouville 
gravity is studied in detail.
Path integral quantisation in generic Euclidean 
dilaton gravity is performed non-perturbatively by analogy to the Minkowskian case.}

\end{titlepage}

\section{Introduction}
Two-dimensional gravities attracted much attention over the last decades
since they retain many properties of their higher-dimensional counterparts
but are considerably more simple. It has been demonstrated that
all two-dimensional dilaton gravities are not only classically
integrable, but this even holds at the (non-perturbative) quantum level (for a recent review cf.~ref.~\cite{Grumiller:2002nm}). However, these
integrability statements have been obtained for the Minkowski 
signature of the space-time only. Many applications require 
Euclidean signature as well. The most prominent example is string
theory where the genus expansion (a typical Euclidean notion)
plays an important role. Spectacular achievements in the Liouville 
model \cite{Nakayama:2004vk}, where higher order correlation functions
have been calculated and relations to matrix models have been established,
are the main motivation for our present work.

The prime goal of this paper is to show classical and quantum integrability
of two-dimensional dilaton gravities in Euclidean space. Our main example
will be (generalised) Liouville gravity, which will be studied in 
full detail locally and globally.

We start with classical solutions. For a generic model all local solutions 
are obtained exploiting the complexified first order formalism and some 
general statements are made regarding the global structure (sect.~\ref{se:2}). 
Then we turn to Liouville gravity, which we re-derive from the
bosonic string sigma model (sect.~\ref{se:3}). A peculiar feature of
our approach is that we do not fix the conformal gauge and keep all
components of the metric dynamical, as well as the Liouville field
(which we call the dilaton, to keep the terminology consistent
with other 2D models). For a three-parameter family of Liouville type
gravities we construct all local and global solutions (sect.~\ref{sLi}). 
Quantisation of Euclidean models (sect.~\ref{se:5}) appears to be somewhat
more complicated than the one of their Minkowski signature counterparts
(which can be traced to the absence of the light-cone condition
in the Euclidean space). Nevertheless, an exact non-perturbative path
integral is performed and local quantum triviality is demonstrated. 
However, this does not mean that all correlators
are trivial. As an illustration, we calculate some non-local
correlation functions. In sect.~\ref{se:conclu} we briefly discuss
perspectives of our approach.
Appendix \ref{AppA} recalls the equivalence between second and first order formulations for Euclidean signature while appendix 
\ref{AppB} considers finite volume corrections to the Liouville action.

\section{Classical solutions}\label{se:2}
\subsection{The model}
The action of general dilaton gravity \cite{Russo:1992yg,Odintsov:1991qu}
(cf.~also \cite{Banks:1991mk,Frolov:1992xx,Mann:1993yv})
\begin{equation}
L_{\mathrm{dil}}=\int d^2\sigma \sqrt{g} \left[
-R\frac \Phi{2} + (\nabla \Phi )^2 \frac{U(\Phi )}2 + V(\Phi )\,, \right]
\label{dilact}
\end{equation}
contains two arbitrary functions $U(\Phi )$ and $V(\Phi )$. An eventual
dependence on an additional function $Z(\Phi)$ in the first term can be
eliminated by a field redefinition if $Z$ is invertible.\footnote{If it is not
  invertible generically curvature singularities arise for $Z^\prime=0$. Thus,
  by demanding invertibility we merely exclude cases with further
  singularities leading to global properties which do not show generic
  differences with respect to those to be encoded in $U(\Phi)$ and $V(\Phi)$ \cite{blaschke04}.}

As for Minkowskian signature 
it is convenient to work with the equivalent first order action 
\cite{Klosch:1996fi,Schaller:1994es}
\begin{equation}
L=\int_\mathcal{M} \left[ Y^a De^a + \Phi d\omega 
+\epsilon \mathcal{V} (Y^2,\Phi) \right]\,, \label{act1}
\end{equation}
where $Y^2=Y^aY^a=(Y^1)^2 + (Y^2)^2$, $e^a$ is the Zweibein 1-form, the 1-form $\omega$ is related to the spin-connection by means of $\omega^{ab}=\varepsilon^{ab} \omega$, with the total anti-symmetric $\varepsilon$-symbol, and the torsion 2-form reads explicitly
\begin{equation}
De^a=de^a + \omega^{ab}\wedge e^b\,.\label{not1}
\end{equation}
The volume 2-form is denoted by $\epsilon=-\frac 12 \varepsilon^{ab} e^a\wedge e^b$. One may raise and lower the flat indices $a,b,\dots$ with the flat Euclidean metric $\delta_{ab}=\rm diag(1,1)$, so there is no essential difference between objects with upper indices and objects with lower ones. Therefore, we do not discriminate between them.
For a more detailed explanation of our notation and conventions of signs see appendix \ref{AppA}.
In this paper we mostly use the Cartan formalism of differential forms.
An introduction to this formalism can be found in \cite{Eguchi:1980jx,Ivey:2003},
specific two-dimensional features are explained in \cite{Grumiller:2002nm}.
Different models are encoded by the ``potential'' $\mathcal{V}$. Here, we restrict ourselves to 
\begin{equation}
\mathcal{V}= \frac 12 Y^2 U(\Phi) +V(\Phi)\,, \label{restrict}
\end{equation}
although our analysis can be extended easily to a more general choices. The
action (\ref{act1}) with the potential \eqref{restrict} describes practically all gravity models
in two dimensions, e.g.\ the Witten black hole ($U = - 1/(2 \Phi)$, $V = -
\lambda^2 \Phi$) and the Schwarzschild black hole ($U = - 1/(2 \Phi)$, $V = -
\lambda^2$). Important exceptions are dilaton-shift invariant models
\cite{Grumiller:2002md} inspired by
the so-called exact string black hole \cite{Dijkgraaf:1992ba}, an action of which has been constructed only recently \cite{Grumiller:2005sq}.

Equivalence of second and first order formalisms (with the choice
(\ref{restrict}))
is proved in appendix
\ref{AppA}, where it is also explained how to translate our notations
to the component language.
\subsection{Local analysis}\label{se:2.2}
In terms of complex fields
\begin{align}
 Y&= \frac{1}{\sqrt{2}} \left( Y^1 +i Y^2\right)\,, & \bar{Y} &=
 \frac{1}{\sqrt{2}} \left( Y^1 -i Y^2\right), \nonumber \\
 e&=\frac{1}{\sqrt{2}} \left( e^1 +i e^2\right)\,,
& \bar e&=\frac{1}{\sqrt{2}} \left( e^1 -i e^2\right)\,, \label{cfields}
\end{align}
the action (\ref{act1}) reads
\begin{equation}
L=\int_\mathcal{M} \left[ \bar Y De + Y\overline{De} 
+ \Phi d\omega 
+\epsilon \mathcal{V} (2\bar YY,\Phi) \right]\,, \label{act2}
\end{equation}
where
\begin{equation}
De:=de -i\omega\wedge e\,,\qquad 
\overline{De}:=d\bar e +i \omega \wedge \bar e \,.\label{not2}
\end{equation}
The volume 2-form is defined as
\begin{equation}
\epsilon :=i\bar e\wedge e \,.\label{vform}
\end{equation}

We at first disregard the conditions implied by (\ref{cfields}) and
solve the model for arbitrary complex fields $\bar Y$, $Y$, $\bar e$
and $e$. If we also allow for complex $\omega$,
we may treat Euclidean and Minkowskian theories
simultaneously. In Minkowski space $\{ Y,\,\bar Y,\, e,\, \bar e,\, \omega \}$
should be replaced by $\{ Y^+,\, Y^-,\, ie^+,\, ie^-,\, i\omega\}$ where
all fields $Y^\pm,\, e^\pm,\, \omega$ are real (cf.~\cite{Grumiller:2002nm}).
The superscript $\pm$ denotes the light-cone components. Euclidean signature imposes a different set of reality conditions to be derived below.

Our complex model shares many similarities with complex Ashtekar
gravity in four dimensions \cite{Ashtekar:1991hf} and relates with the so-called
generalised Wick transformation \cite{Ashtekar:1996qw}. A two-dimensional
version (spherical reduction) has been discussed in \cite{Thiemann:1993jj}.
The main difference from usual theories of complex fields (say, from
$|\phi|^4$ theory) is that if no relation between 
$\bar Y$, $Y$, $\bar e$ and $e$ is assumed the action (\ref{act2})
is holomorphic rather than real. Path integration in such theories
requires the contour integration in the complex plane 
rather than the Gaussian integration
\cite{Alexandrov:1998cu}.

Another reason to study complex gravity theories is their
relation to noncommutative models. In that case
the Lorentz group does not close and one has to consider a
complexified version \cite{Chamseddine:2000zu,Moffat:2000gr}.
In two dimensions, a noncommutative version of 
Jackiw-Teitelboim gravity \cite{JT1,JT2,JT3,JT4} was constructed in
\cite{Cacciatori:2002ib} and later quantised in 
\cite{Vassilevich:2004ym}. Gauged noncommutative Wess-Zumino-Witten models have been studied in ref.~\cite{Ghezelbash:2000pz}. Complexified Liouville gravity was shown to be equivalent to an $SL(2,R)/U(1)$ model \cite{Jorjadze:2003zg}.

The equations of motion from variation of $\omega$, $\bar e$, $e$, $\Phi$, $\bar Y$ and $Y$, respectively,
read
\begin{eqnarray}
&&d\Phi-i\bar Y e + i Y\bar e=0 \,,\label{oeq}\\
&&DY+ie \mathcal{V}=0 \,,\label{beeq}\\
&&\overline{DY} -i\bar e \mathcal{V}=0\,,\label{beq}\\
&&d\omega + \epsilon \frac{\partial \mathcal{V}}{\partial \Phi} =0
\,,\label{Xeq}\\
&&De 
+ \epsilon \frac{\partial \mathcal{V}}{\partial \bar Y} =0\,,\label{bYeq}\\
&&\overline{De}
+ \epsilon \frac{\partial \mathcal{V}}{\partial Y} =0\,.\label{Yeq}
\end{eqnarray}
To obtain the solution of \eqref{oeq}-\eqref{Yeq} we follow the method of \cite{Grumiller:2002nm} 
(cf.\ \cite{Kummer:1995qv} and for a different approach \cite{Schaller:1994np}). 
By using (\ref{oeq}) - (\ref{beq})
it is easy to demonstrate that
\begin{equation}
d(\bar Y Y)+\mathcal{V}d\Phi=0 \,.\label{gencons}
\end{equation}
Noting that $Y^2/2=\bar{Y}Y$ with the choice (\ref{restrict}) for
$\mathcal{V}$ one can integrate (\ref{gencons})
to obtain the local conservation law
\begin{equation}
d\mathcal{C}=0,\qquad \mathcal{C}=w(\Phi) +e^{Q(\Phi)} \bar Y Y \label{C}
\end{equation}
with
\begin{equation}
Q(\Phi)=\int^\Phi U(y) dy\,,\qquad w(\Phi)=\int^\Phi e^{Q(y)}V(y)dy \,.\label{Qw}
\end{equation}
Next we use (\ref{beeq}) to express
\begin{equation}
\omega = Z \mathcal{V} -i\frac{dY}Y \,,\label{oZ}
\end{equation}
where
\begin{equation}
Z:=\frac 1Y e \,.\label{defZ}
\end{equation}
We have to assume\footnote{For Euclidean signature $Y=0$ implies $\bar{Y}=0$. Solutions where
$Y=0=\bar{Y}$ globally may be analysed separately and admit only constant
curvature. $Y$ and $\bar{Y}$ cannot vanish at isolated points inside the manifold.}
that $Y\ne 0$. Now, by means of (\ref{oeq}) we can show
that
\begin{equation}
\bar e=i\frac {d\Phi}Y +Z{\bar Y} \,,\label{beZ}
\end{equation}
and with \eqref{bYeq} that
\begin{equation}
dZ=d\Phi\wedge Z U(\Phi) \,.\label{dZZ}
\end{equation}
Equation (\ref{dZZ}) yields
\begin{equation}
d\hat Z = 0\,,\qquad \hat Z:=e^{-Q} Z \,.\label{ZhZ}
\end{equation}
This equation can be integrated locally\footnote{Globally (\ref{hZf}) is true
up to a harmonic one-form (see sec.\ \ref{sGS}).}:
\begin{equation}
\hat Z=df \,,\label{hZf}
\end{equation}
where $f$ is a complex valued zero-form. Besides \eqref{C} this is the only integration
needed to obtain the solution of the model, which depends on two arbitrary complex functions,
$f$ and $Y$, one arbitrary real function $\Phi$, and one constant
of motion $\mathcal{C}$. $\bar Y$ is then defined by the conservation
law (\ref{C}), and the geometry is determined by
\begin{eqnarray}
&&e=Ye^{Q}df \,,\label{sole}\\
&&\bar e=i\frac{d\Phi}Y+\bar Y e^Q df \,,\label{solbe}\\
&&\omega = \mathcal{V} e^Q df -i\frac {dY}Y \,.\label{solom}
\end{eqnarray}
One can check easily that indeed all the equations (\ref{oeq}) -
(\ref{Yeq}) are satisfied. This completes the local classical
analysis of complex dilaton gravity in 2D.

To proceed with real
Euclidean dilaton gravity the two reality conditions are needed which follow
from \eqref{cfields}:
\begin{eqnarray}
&&\bar Y=Y^* \,,\label{1rc}\\
&&\bar e=e^* \,.\label{2rc}
\end{eqnarray}
Here star denotes complex conjugation.
The first of these conditions implies that $\mathcal{C}$ is real.
The second one fixes the imaginary part of $Z$:
\begin{equation}
\Im Z=-\frac 12 \frac {d\Phi}{\bar YY} \,.\label{ImZ}
\end{equation}
The line element 
\begin{equation}
(ds)^2=2\bar e e=2\bar YYe^{2Q} (d\theta )^2 + \frac{(d\Phi)^2}{2\bar YY}
\label{ds21}
\end{equation}
(with $\theta :=\Re f$) is now explicitly real. The product $\bar YY$ can
be eliminated by means of the conservation law (\ref{C}) so that
\begin{equation}
(ds)^2=e^{Q(\Phi)} \left( \xi (\Phi) (d\theta )^2 
+ \xi(\Phi)^{-1} (d\Phi)^2 \right),
\quad \xi=2(\mathcal{C}-w(\Phi))\,. \label{ds22}
\end{equation} 
This solution always has a Killing vector $\partial_\theta$ and depends on one
constant of motion $\mathcal{C}$.
The Levi-Civita connection $\hat \omega$
and the scalar curvature corresponding to the line element (\ref{ds22})
read
\begin{equation}
\hat\omega = \frac 12 (U\xi + \xi')d\theta \,,\qquad
R=-e^{-Q} (U'\xi + U\xi' + \xi'') \,, \label{hoR}
\end{equation}
where prime denotes differentiation w.r.t.~$\Phi$. 

It should be noted that whereas \eqref{sole} and \eqref{solbe} would
lead to an Eddington-Finkelstein type line element in complex gravity,
reminiscent of the one in Minkowski space \cite{Grumiller:2002nm}, the reality
conditions \eqref{1rc} and \eqref{2rc} force a gauge of diagonal type. This is
the source of many technical complications for Euclidean signature.
\subsection{Global structure}\label{sGS}
Our strategy is borrowed from Lorentzian signature gravity theories.
We take local solutions (\ref{ds22}) and extend them as far as
possible along geodesics. In this way we obtain all global solutions. Here we only sketch
the methods, postponing a more detailed discussion to the next section
where we treat Liouville gravity as a particular example. 

In the $\Phi$ direction each solution can be extended as long as the metric
is positive definite, $\xi >0$, and as long as it does not hit
a curvature singularity. Note that $\Phi$ plays a dual role. It is not only
one of the coordinates, it is also a scalar field. Therefore, one cannot
identify regions with different values of $\Phi$ as it would lead to
a discontinuity. In the $\theta$ direction one can either extend the solution
to an infinite interval or impose periodicity conditions. It is easy to
see that a smooth solution can be obtained only if the period does not
depend on $\Phi$. This period should be chosen such that
conical singularities are avoided at the roots of $\xi (\Phi)$. In some cases one
can also ``compactify'' the manifold by adding a point with $\Phi =\infty$
(see below). 

From a different point of view the action (\ref{act1}) can be considered as a particular case of a Poisson
Sigma Model (PSM) \cite{Schaller:1994es,Ikeda:1994fh}. Classical solutions of PSMs on Riemann
surfaces have been analysed recently in \cite{Bojowald:2003pz}. It has
been found that on an arbitrary surface one encounters a 
$(1+\mathrm{dim}\, H^1(\mathcal{M}))$-parametric family of solutions.
Roughly speaking, one of the parameters is the conserved quantity 
$\mathcal{C}$. The others originate from arbitrariness in global
solutions of (\ref{ZhZ}) which is exactly the dimension of the first
cohomology group of the underlying manifold $\mathcal{M}$.

Here we use a different approach. In a gravity theory, the topology of the
manifold has to be consistent with the metric.\footnote{An example of restrictions imposed on topology by the Riemannian 
structure is the Gauss-Bonnet density, the integral of which 
essentially provides the Euler characteristic.} For example,
solutions with $\xi <0$ are perfectly admissible for a PSM, but make
no sense in a Euclidean gravity theory. Also, curvature singularities play
an important role in analysing the global structure  of classical solutions
in gravity, but are not discussed in the PSM approach
\cite{Bojowald:2003pz}.

In the particular case of the Katanaev--Volovich model
\cite{Katanaev:1986wk,Katanaev:1990qm}
a complete analysis of local and global solutions in the conformal gauge
has been performed in ref.~\cite{Katanaev:1997je}. 

\section{Liouville gravity from bosonic strings}\label{se:3}
Before discussing in detail classical solutions of the Liouville gravity
model, we re-derive the Liouville action from string theory. Our
derivation does not imply conformal gauge fixing and, therefore,
differs from more standard ones (cf.~the discussion at the end of
this section).

The starting point is the bosonic string partition function
\begin{equation}
\mathcal{Z}(l)=\int [\mathcal{D}g]\, \mathcal{D} X
\exp\left( -L(X,g) -\mu_0 \int d^2\sigma \sqrt{g} 
+ \int d^2\sigma l_A X^A\right)\,,\label{partf}
\end{equation} 
where the ``matter part'' corresponds to the string action
\begin{equation}
L(X,g)=\frac 12 \int d^2\sigma \sqrt{g} g^{\mu\nu} \partial_\mu X^A
\partial_\nu X_A \,.\label{mact}
\end{equation}
Here $X$ is a scalar field with values in $\mathbb{R}^N$. 
We have introduced sources $l_A$ for $X^A$, which are
scalar densities from the point of view of the world sheet. 
We assume
Euclidean signature on the target space and on the world sheet. The measure
$[\mathcal{D}g]$ includes all usual gauge parts (ghosts and gauge fixing).

One can integrate over $X$ by using a procedure suggested by Polyakov
\cite{Polyakov:1981rd} (cf.\ also an earlier paper
\cite{Luscher:1980fr}). The effective action $W(g,l)$ is defined by
\begin{eqnarray}
W(g,l)&&=-\ln \int \mathcal{D}X \, e^{-L(X,g)+\int d^2\sigma l_A X^A} 
+ \mu_0 \int d^2\sigma \sqrt{g} \nonumber \\
&&=\frac N2 \ln \det (-\Delta) +\frac 12 \int d^2\sigma \sqrt{g} l \Delta^{-1} l
+ \mu_0 \int d^2\sigma \sqrt{g} 
\,,\label{Wg}
\end{eqnarray}
where $\Delta=g^{\mu\nu}\nabla_\mu\nabla_\nu$. The presence of
$l$ allows us to calculate correlation functions of $X$ although
this field has been integrated out already. However, from now on we put $l=0$
and define $W(g)=W(g,0)$. 

We wish to retain one out of the $N$ contributions in the first term on the
right hand side of (\ref{Wg}). The remaining $N-1$ terms can be fixed
by the conformal
anomaly
\begin{equation}
T_\mu^\mu = \frac {N-1}{24\pi} [ R + \tilde \mu ] \,,\label{Tmm}
\end{equation}
where $T_{\mu\nu}$ is the vacuum expectation value of
the energy-momentum tensor for $N-1$ fields $X$. 
$R$ is the Riemann curvature
of the two-dimensional metric $g_{\mu\nu}$. 
The term with $\tilde\mu$ describes renormalisation 
of the cosmological constant. It is 
``non-universal'', i.e. it does not appear in certain regularisations, as, 
e.g., in the zeta function one.

Using the well-known relation between $T_\mu^\mu$ and the effective action in
the terms $\propto (N-1)$ yields\footnote{On a compact manifold the operator $\Delta$ has zero modes. Therefore,
the argument has to be modified (cf.\ 
Appendix \ref{AppB}).}
\begin{equation}
W(g)=\frac {N-1}{96\pi}  \int d^2\sigma \sqrt{g} \left[ R \frac 1{\Delta} R
+ \mu \right]+\frac 12 \ln \det (-\Delta) \,,\label{nonlocP}
\end{equation}
where the constant $\mu$ contains contributions from $\mu_0$ and $\tilde\mu$.

This action can be transformed to a local one by introducing an additional
scalar field $\Phi$,
\begin{equation}
W(g,\Phi) = \frac {N-1}{24\pi}  \int d^2\sigma \sqrt{g} \left[
(\partial \Phi)^2 -\Phi R +\mu /4\right]\,, \label{locP}
\end{equation} 
so that the partition function (\ref{partf}) becomes
\begin{equation}
\mathcal{Z}=\int [\mathcal{D}g]\, \mathcal{D} \Phi e^{-W(g,\Phi)}\,.
\label{Zgr}
\end{equation}
The integration over $\Phi$ produces both the non-local term and
the $\ln \det (-\Delta)$ in (\ref{nonlocP}) which had been separated there
just for this purpose. One could perform a
rescaling $\Phi \to \sqrt{N-1}\Phi$. Then the limit $N\to 1$
is consistent (but not particularly interesting) 
as it reproduces our starting point (\ref{partf})
without external sources.

An exponential interaction term can be obtained by a conformal 
transformation of the metric:
\begin{equation}
W(e^{\alpha\Phi}\tilde g, \Phi)=\frac {N-1}{24\pi}  
\int d^2\sigma \sqrt{\tilde g} \left[
(1-\alpha)(\partial \Phi)^2 -\Phi\tilde R +\mu e^{\alpha\Phi} /4\right]
\,, \label{expint}
\end{equation} 
where $\alpha$ is a constant. Note that the conformal transformation $g\to\tilde g$ is singular for $\Phi\to\pm\infty$.

It is instructive to compare our approach to the one of Ref.\ \cite{David:1988hj,Distler:1989jt},
which also derives the Liouville action from quantum strings. 
There this action is obtained from
fixing the conformal gauge while we integrate over the string coordinates
$X^A$. As a result, the action of Refs.\ \cite{David:1988hj,Distler:1989jt} contains somewhat
different numerical coefficients, and, more important, the metric
there is non-dynamical. If one keeps the sources $l^A$ for $X^A$, both
approaches should give equivalent quantum theories. As we shall see below,
our results for geometries of ``Liouville gravity'' are consistent with
the semiclassical results in the conformal gauge approach (cf.\ e.g.\ \cite{Ginsparg:1993is}).

\section{Classical solutions of Liouville gravity}\label{sLi}

In accordance with (\ref{expint}) we restrict ourselves to the first order action (\ref{act1}) with a potential of type (\ref{restrict}) where
\begin{equation}
U(\Phi)=a\,,\qquad V(\Phi)=be^{\alpha\Phi}\,,\label{UVLi}
\end{equation}
with $a,b,\alpha\in \mathbb{R}$. Note that one of the
constants can be absorbed in a redefinition of $\Phi$.
From (\ref{Qw}) one immediately finds that for $(a+\alpha)\neq 0$
\begin{equation}
Q(\Phi)=a\Phi, \qquad w(\Phi)=\frac{b}{a+\alpha} e^{(a+\alpha)\Phi}
\,.\label{QwLi}
\end{equation}
The simple case $a+\alpha=0$ yields $w=b\Phi$ and will be treated separately. Equation (\ref{ds22}) becomes
\begin{eqnarray}
&&(ds)^2=e^{a\Phi} \left( \xi (\Phi) (d\theta )^2 
+ \xi(\Phi)^{-1} (d\Phi)^2 \right),\nonumber\\
&& \xi=2\left( \mathcal{C}-\frac{b}{a+\alpha} e^{(a+\alpha)\Phi}
\right)\,, \label{dsLi}
\end{eqnarray}
yielding the scalar curvature:
\begin{equation}
R=2be^{\alpha\Phi} (2a+\alpha) \,.\label{RLi}
\end{equation}
It should be emphasised that $R$ is independent of $\mathcal{C}$, a feature
which is {\em not} true generically but holds if and only if $U=\rm const.$,
as seen from (\ref{hoR}).
At $\alpha\Phi \to -\infty$ we have an asymptotically flat region;
$\alpha\Phi \to +\infty$ corresponds to a curvature singularity.
At $\Phi=\Phi_0$ such that $\xi (\Phi_0)=0$ there is a coordinate
singularity. In a space with Minkowski signature $\Phi=\Phi_0$ corresponds
to a Killing horizon.

Depending on the sign of $a$, $b$ and $\alpha$ different global geometries
are possible. We consider them case by case.

\subsection{Generic solutions}\label{sec:laulimamelina}

First we assume $a>0$. The case $a<0$ can be obtained by changing the sign of $\Phi$. The limit $a=0$ is addressed in the subsequent subsection. 

\paragraph{Case I:} $b,\alpha > 0$. 

For $\mathcal{C}\le 0$ the metric (\ref{dsLi}) is negative definite,
so that no solution in Euclidean space exists. For $\mathcal{C}> 0$,
the coordinate $\Phi$ is restricted by the inequality
\begin{equation}
\Phi \le \Phi_0 = \frac 1{a+\alpha} \ln \left(
\frac{ (a+\alpha)\mathcal{C}}b \right) \,,\label{ineq1}
\end{equation}
so that the solution includes the asymptotically flat region but
does not contain any curvature singularity. 

Near $\Phi_0$ the line element (\ref{dsLi}) becomes
\begin{equation}
(ds)^2\approx e^{a\Phi_0} \left( 2(a+\alpha)\mathcal{C} \phi  (d\theta)^2+
\frac{(d\phi)^2}{2 (a+\alpha)\mathcal{C}\phi} \right) \,,\label{near}
\end{equation}
where $\phi :=\Phi -\Phi_0$. To avoid a conical singularity at $\phi=0$
the coordinate $\theta$ should be periodic with the period
\begin{equation}
\theta_0=\frac{2\pi}{(a+\alpha)\mathcal{C}} \,.\label{period}
\end{equation}
We see that these solutions correspond to a characteristic temperature
which, however, cannot be identified directly with $1/\theta_0$ since the
metric (\ref{dsLi}) is not explicitly the unit one in the
flat region $\Phi\to -\infty$. 

Although the coordinate $\Phi$ varies from $-\infty$ to
$\Phi_0$, the proper geodesic distance along the line $\theta=const.$
between $\Phi=-\infty$ and $\Phi=\Phi_0$ is finite (and easy to calculate
but looks ugly). This suggests that $\Phi=-\infty$ is a boundary, and
we have a manifold with the topology of a disc. To make sure that this
is really the case, let us calculate the Euler characteristic of the
manifold. The volume contribution
\begin{equation}
\chi_{\mathrm{vol}}=\frac 1{4\pi} \int_{\mathcal{M}} R \sqrt{g} d^2\sigma
= \frac{2a+\alpha}{a+\alpha} \label{volEu}
\end{equation}
is not an integer. However, if we add the boundary part
\begin{equation}
\chi_{\mathrm{bou}}=\frac 1{2\pi} \int_{\Phi=-\infty} k d\tau =
-\frac{a}{a+\alpha} \label{bouEu}
\end{equation}
(where $k$ is the trace of the second fundamental form of the boundary,
and $d\tau$ is the arc length along the boundary), we obtain
\begin{equation}
\chi=\chi_{\mathrm{vol}}+\chi_{\mathrm{bou}}=1 \,,\label{totEu}
\end{equation}
which coincides with the Euler characteristic of the disc.

This metric exhibits a somewhat strange behaviour: the radius
of the circle $\Phi =const$ goes to zero as $\Phi\to -\infty$.
Therefore, it seems natural to add that point to the
manifold. This has to be done smoothly. To avoid a conical singularity
there the period of the $\theta$ coordinate should be
\begin{equation}
\tilde\theta_0=\frac{2\pi}{\mathcal{C}a} \,.\label{ttwid}
\end{equation}
A smooth manifold can only be achieved if $\tilde\theta_0=\theta_0$
yielding $\alpha =0$. This special case is considered below (Case sI).

Adding a point to the manifold is only possible at $\Phi =\pm\infty$.
The point we add should be at a finite distance from the points inside
the manifold, should correspond to a region with a finite curvature,
and the circles $\Phi =const$ should have zero radius at this point.
All these restrictions can be satisfied for the cases I (above) and
sI (below).

\paragraph{Case II:} $\alpha > 0$, $b<0$.

Here both signs of the conserved quantity $\mathcal{C}$ are
possible. For $\mathcal{C}<0$ the coordinate $\Phi$ ranges from 
$\Phi_0$ to $+\infty$ (curvature singularity). The conical
singularity at $\Phi=\Phi_0$ is avoided by imposing the same periodicity 
condition (\ref{period}) on $\theta$.
The geodesic distance between $\Phi=\Phi_0$ and $\Phi=+\infty$
is finite. 

For $\mathcal{C}>0$ there is no point such that $\xi (\Phi)=0$. Therefore,
the period in $\theta$ is not fixed and $\Phi \in [-\infty,\infty[$. The flat
region ($\Phi =-\infty$) is separated by a finite
distance from the curvature singularity ($\Phi=+\infty$). Since all
solutions contain a singularity this case does not fit into the discussion of
standard topologies.



\paragraph{Case III:} $\alpha < 0$, $a+\alpha >0$.

For $b>0$ solutions with $\mathcal{C}<0$ are excluded. If $\mathcal{C}>0$,
then $\Phi \in ]-\infty,\Phi_0]$, i.e. all solutions contain a curvature
singularity and have to be periodic. The distance to the singularity is
finite. 

For $b<0$ and $\mathcal{C}<0$ we encounter non-compact periodic solutions
with $\Phi \in [\Phi_0,+\infty[$, which includes the flat region but not
the singularity. The total distance along the geodesic $\theta=const$
is infinite. This is a deformed Euclidean plane. 
The situation for $\mathcal{C}>0$ is similar, but then
all solutions also include the curvature singularity.

\paragraph{Case IV:} $a>0$, $\alpha < 0$, $a+\alpha <0$.

This case can be easily analysed along the same lines, exchanging $b\to-b$.



\subsection{Special solutions}

\paragraph{Case sI:} $\alpha=0$, $a,b\neq 0$

This is a rather special\footnote{%
 It is evident from \eqref{RLi} that curvature is constant for $\alpha=0$, just
 like for the Jackiw-Teitelboim model \cite{JT1,JT2,JT3,JT4}. Nevertheless, the resulting
 action is inequivalent to the one of that model because both potentials
 \eqref{UVLi} differ from $U_{\rm JT}=0$, $V_{\rm JT}\propto\Phi$. It is not even
 conformally related because the conformally invariant function $w$ in \eqref{QwLi}
 is not quadratic in the dilaton $\Phi$, as required for the
 Jackiw-Teitelboim model and its conformally related cousins.} 
but the most important particular case since it
corresponds to the original action (\ref{locP}) before the conformal
transformation. Obviously, the sign of $a$ is not significant as it can be
reabsorbed into a reflection of $\Phi$ and $b$. We therefore
take $a>0$. For $b>0$ negative values of $\mathcal{C}$ are not allowed.
For positive $\mathcal{C}$ we encounter a manifold with constant positive
scalar curvature which can be identified with a two-sphere $S^2$
after adding a point corresponding to $\Phi =-\infty$. Adding this point
removes the boundary contribution from (\ref{totEu}) and for $\alpha=0$ Eq.\ (\ref{volEu})  establishes $\chi=2$ as required for $S^2$. For $b<0$ we have
hyperbolic spaces with negative constant scalar curvature. For
$\mathcal{C}>0$ the solution\footnote{This case corresponds to the Katanaev-Volovich model \cite{Katanaev:1986wk,Katanaev:1990qm} for vanishing $R^2$ term, i.e., a theory with constant torsion and cosmological constant. Its Euclidean formulation has been discussed in ref.~\cite{Schaller:1993ei}.} should be completed by $\Phi=+\infty$
and $\theta$ should be taken periodic with the period (\ref{ttwid}).

\paragraph{Case sII:} $a=0$, $\alpha,b\neq 0$

Taking the formal limit $a\to 0$ in (\ref{bouEu}) and (\ref{volEu}) provides
$\chi=1$, i.e.~the result for a disc. However, depending on the signs of
$\alpha$ and $b$ singularities may be encountered and the manifold need not be compact. In fact, the discussion is in full analogy to the one for generic solutions.

\paragraph{Case sIII:} $a+\alpha=0$, $b\neq 0$

Instead of the second equation in (\ref{QwLi}) one obtains $w=b\Phi$ and the
curvature becomes $R=2abe^{-a\Phi}$. Thus one arrives at a conformally transformed version of the Witten black hole \cite{Mandal:1991tz,Elitzur:1991cb,Witten:1991yr} with topology of a ``cigar''. A conformal frame that is often employed \cite{Callan:1992rs} exploits the simplicity of the geometry for $a=0$: the metric is flat.

\paragraph{Case sIV:} $b=0$

Here the value of $\alpha$ is irrelevant and the solutions are flat. The
period reads $\theta_0=2\pi/(\mathcal{C}a)$. If additionally $a=0$ both potentials vanish and toric topology is possible by identifying $\Phi$ periodically. Then the Euler characteristic vanishes trivially. This is the only case where the Euler characteristic is not positive.

\subsection{Concluding remarks}

It is a common feature of all finite volume solutions that they
are periodic in $\theta$ and do not include the region $a\Phi \to +\infty$.
This completes our analysis of classical solutions in Liouville
gravity.

A final remark in this section regards the Euler characteristic which
worked so well in the Case I above. To define $\chi$ for a non-compact
manifold, it has to be placed into a box. Otherwise, the curvature integral
yields an incorrect (infinite or non-integer) value which is not
related to the index \cite{Eguchi:1980jx}. 

\section{Quantisation}\label{se:5}

The quantisation for Minkowskian signature in the absence of matter has been
performed for various dilaton gravity models in different formalisms. It is
not necessary to review all these approaches and results (cf.~\cite{Grumiller:2002nm} and references therein, especially sections 
7.-9.; see also \cite{Louis-Martinez:1994eh}). Here we
follow the path integral quantisation
\cite{Haider:1994cw,Kummer:1997hy}. 

Employing the complex formalism of sect.~\ref{se:2.2} there are two possible
strategies: either to quantise first and to impose the reality condition
afterwards by hand or the other way round. Clearly, the second route is the
one that {\em should} (and will be) be pursued\footnote{It should be noted that  the path integral quantisation 
of complex gravity \cite{Alexandrov:1998cu} is based upon
equivalence to a real version.}. 
Alas, at first glance it appears that it is only the first one that {\em can} be pursued. In the following, we will discuss why and how.

The crucial technical ingredient which makes all path integrations for
Min\-kow\-skian signature straightforward was to impose a temporal gauge for light-like Cartan variables $e^{\pm}$, which corresponds to the
Eddington-Finkelstein gauge in the metric
\cite{Haider:1994cw,Kummer:1997hy}. This is quite easy to understand: the
action of type \eqref{act1} is bilinear in the gauge fields $\omega$ and
$e^a$. If e.g.\ the components $\omega_1$, $e_1^a$ are gauged to constants
then the gauge fixed action becomes linear in the remaining field
components. Provided that no complications arise from the ghost sector (which
indeed happens to be true) there will be as many functional $\delta$-functions
as there are gauge fields, which in turn can be used to integrate out the
target space coordinates $\Phi$, $Y^a$. As a result the quantum effective action is nothing but the {\em classical} action, up to boundary contributions. 

\subsection{Gauge fixing and path integral}

The concept of the Eddington-Finkelstein gauge is intimately related to the Minkowskian signature, but it can be generalised readily to complex fields, e.g.
\begin{equation}
  \label{eq:ef}
  e_1=0\,,\quad\bar{e}_1=c\,,\quad\omega_1=0\,,
\end{equation}
with some non-vanishing $c\in\mathbb{C}$. Thus, the program as outlined above
can be applied to quantise complex dilaton gravity in the first order
formalism. Again, local quantum triviality can be established and the quantum
effective action equals the classical one, up to boundary contributions. One
can now try to impose the reality conditions (\ref{1rc}), (\ref{2rc}) on the
mean fields appearing in that action to obtain Euclidean signature. However,
on general grounds it is clear that imposing such conditions and quantisation
need not commute; apart from that, the reality condition
(\ref{2rc}) is incompatible with the gauge choice (\ref{eq:ef}) for reasons
described below. Thus, if one is interested in rigorous quantisation of
Euclidean dilaton gravity one ought to impose the reality conditions
(\ref{1rc}), (\ref{2rc}) {\em first}. This leads to a serious problem, namely
that the notion of ``light-like'' field components ceases to make sense and
the gauge (\ref{eq:ef}) no longer is applicable. The fastest way to check this
statement is to recognise that (\ref{2rc}) implies $c=0$ in (\ref{eq:ef})
which amounts to a singular gauge with degenerate metric. In other words, such a choice is not accessible for Euclidean reality conditions and has to be discarded. 

These difficulties can be avoided if instead of (\ref{eq:ef}) we employ
\begin{equation}
  \label{eq:efeuk}
  e_1=c\,,\quad\bar{e}_1=c^\ast\,,\quad\omega_1=0\,,
\end{equation}
with some non-vanishing $c\in\mathbb{C}$. Such a gauge is compatible with the reality condition (\ref{2rc}). The metric becomes
\begin{equation}
  \label{eq:metric}
  g_{\mu\nu}=\left(\begin{array}{cc}
  2|c|^2 & ce_2^\ast+c^\ast e_2\\
  ce_2^\ast+c^\ast e_2 & 2|e_2|^2
  \end{array}\right)\,.
\end{equation}

The path integral quantisation based upon the formulation \eqref{act1} now can be performed in full analogy to 
\cite{Kummer:1997hy} (see also section 7 of \cite{Grumiller:2002nm}). 
As there arise no subtleties in the ghost and gauge fixing sector after
(trivially) integrating out $e_1$, $e_1^*$, $\omega_1$ and the ghosts
we only obtain a contribution $\mathcal{F}$ to the Faddeev-Popov determinant:
\begin{equation}
\mathcal{F}=\det 
\begin{pmatrix}
  -\partial_1 & - i c^\ast \mathcal{V}' & i c \mathcal{V}' \\
  i c^\ast & -i c^\ast Y^\ast \dot{\mathcal{V}} &  -\partial_1 + i c Y^\ast
  \dot{\mathcal{V}} \\
  - ic & -\partial_1-ic^\ast Y\dot{\mathcal{V}} &  icY\dot{\mathcal{V}}
\end{pmatrix}
\label{FPdet}\end{equation}
Here $\mathcal{V}' = \partial_\Phi \mathcal{V}$ and $\dot{\mathcal{V}} =
\partial_{YY^\ast} \mathcal{V}$. 
This result can be checked easily by using a gauge fixing fermion implying 
a ``multiplier gauge'' \cite{Grumiller:2001ea}. Then, the 
matrix $\mathcal{F}$ is given by the ``structure functions'' in full analogy to the 
Minkowskian case (cf.~Eqs.~(E.117) and (E.118) of \cite{Grumiller:2001ea}). 
The only difference is that instead of containing a single ghost-momentum 
the gauge fixing fermion now contains two ghost-momenta, multiplied by 
$c,c^\ast$, respectively. Consequently, also $\mathcal{F}$ 
consists of two terms in such a way that a real expression emerges. 
The crucial observation is that the determinant \eqref{FPdet} is independent 
from $\omega$ and $e^\pm$. 

After this step the generating functional for the Green functions
reads
\begin{multline}
\mathcal{Z}(j,J)=\int \mathcal{D}e_2  \mathcal{D}e_2^*\mathcal{D}\omega_2
\mathcal{D}Y \mathcal{D}Y^*\mathcal{D}\Phi\, \mathcal{F}\, \\
\times\exp\left[ i\int d^2x \left(
L_{\rm g.f} 
+je_2 +\bar j e_2^* +j_\omega \omega_2 
+JY+\bar J Y^*+J_\Phi \Phi \right) \right]\,,
\label{ZjJ}
\end{multline}
where $j$, $J$ are external sources. Reality implies 
$j_\omega\in\mathbb{R}$ and $\bar{j}=j^\ast$. As a result, the classical action has been replaced by its
gauge-fixed version
\begin{equation}
  \label{eq:gfa}
  L_{\rm g.f.} = Y^\ast(\partial_1e_2+ic\omega_2)
+Y(\partial_1e_2^\ast-ic^\ast\omega_2)
+\Phi\partial_1\omega_2+i(c^\ast e_2-ce_2^\ast){\mathcal V}\,.
\end{equation}
Note that the reality conditions (\ref{1rc}), (\ref{2rc}) already have been imposed in (\ref{eq:gfa}). However, one can undo this restriction simply by replacing all quantities with $\ast$ by barred quantities. 

Due to the linearity of 
(\ref{eq:gfa}) in $e_2$, $e_2^*$ and $\omega_2$, and due to the independence
of (\ref{FPdet}) on these variables
the path integration over those components produces three functional $\delta$ functions with arguments
\begin{align}
& -\partial_1\Phi+icY^\ast-ic^\ast Y+j_\omega \,,\\
& -\partial_1Y^\ast+ic^\ast{\mathcal V}(2YY^\ast,\Phi)+\bar{j} \,,\\
& -\partial_1Y-ic{\mathcal V}(2YY^\ast,\Phi)+j \,.
\end{align}
 It is useful to introduce the real fields
\begin{equation}
  \label{eq:redef}
  Y_1:=c^\ast Y+cY^\ast\,,\quad Y_2:=i\left(cY^\ast-c^\ast Y\right)\,.
\end{equation}
This redefinition yields a constant (nonsingular) Jacobian, as well as $2YY^\ast=(Y_1^2+Y_2^2)/(2|c|^2)$. Integrating out the target space variables with (\ref{restrict}) establishes
\begin{align}
& \partial_1\Phi = Y_2+j_\omega\,, \label{eq:eom1}\\
& \partial_1Y_1= j_1\,, \label{eq:eom2}\\
& \partial_1Y_2=-\frac12 (Y_1^2+Y_2^2)U(\Phi)-2|c|^2 V(\Phi) + j_2\,, \label{eq:eom3}
\end{align}
where $j_1:=cj^\ast+c^\ast j$ and $j_2:=i(cj^\ast-c^\ast j)$.
The inverse determinant which appears due to this integration contains
the same combination of structure functions appearing in the 
matrix (\ref{FPdet}).
Thus, as in the Minkowskian case, there is no Faddeev-Popov determinant left after integrating over {\em all} variables.
For generic gauges an analogous statement can be found in ref.~\cite{Katanaev:2000kc}.

In the absence of sources eq.\ \eqref{eq:eom2} can be integrated immediately,
while eqs.\ \eqref{eq:eom1} and \eqref{eq:eom3} are coupled. Nevertheless, by taking a proper
combination of them it is possible to reproduce the conservation equation
(\ref{gencons}), which then can be integrated as in (\ref{C}). This equation
can be used to express $Y_2$ in terms of $Y_1$ and of $\Phi$. Then, only the
first order equation  (\ref{eq:eom1}) has to be solved. Including source
terms\footnote{For simplicity $j_\omega$ has been set to zero which is
  sufficient to describe situations where only the dependence on the metric is
  important. For $j_\omega\ne 0$ it is still possible to solve (\ref{eq:sol2}) (which remains unchanged) and (\ref{eq:sol3}), which receives additional contributions non-linear in $j_\omega$. These terms have to be added on the r.h.s. of (\ref{eq:sol3}), but the modified version of (\ref{eq:sol3}) still is quadratic (and algebraic) in $Y_2$.} $j_1,j_2$ the general solution for $Y_1$, $Y_2$ can be represented as
\begin{align}
& Y_1= Y_1^h+\partial_1^{-1}j_1\,, \qquad Y_1^h\in\mathbb{R}\,,\label{eq:sol2}\\
& (Y_2)^2=4|c|^2e^{-Q(\Phi)}\left({\mathcal C}_0-w(\Phi)\right)-(Y_1)^2+2\hat{T}\left(j_2\partial_1\Phi+Y_1j_1\right)\,, \label{eq:sol3}
\end{align}
with
\begin{equation}
\hat{T}(f(x^1)):=e^{-Q(\Phi)}\partial_1^{-1}\left(e^{Q(\Phi)}f(x^1)\right)\,.
\end{equation} 
Inserting one of the square-roots of (\ref{eq:sol3}) into (\ref{eq:eom1})
yields a non-linear equation for $\Phi$ which reduces to a first order
differential equation in the absence of sources. Note that the integration
functions $Y_1^h$ and ${\mathcal C}_0$ are not only independent from $x^1$ but
show also independence\footnote{\label{fn:6} It should be emphasised that in addition to (\ref{eq:eom1})-(\ref{eq:eom3}) there are the so-called ``lost constraints'' \cite{Kummer:1998zs,Grumiller:2001ea}, which play the role of gauge Ward identities and
are nothing but the classical equations of motion analogous to (\ref{eq:eom1})-(\ref{eq:eom3}) but with $\partial_1$ replaced by $\partial_2$. Moreover no additional integration constants arise from the equations of motion for the gauge fields if these ``lost constraints'' are taken into account \cite{Grumiller:2002dm}.} from $x^2$. 
In the absence of sources the Casimir function ${\mathcal C}$ coincides
on-shell with the constant ${\mathcal C}_0$. But even when sources are present
the quantity ${\mathcal C}_0$ remains a free constant and therefore fixing
it to a certain value can be used to define the physical vacuum.

To summarise, the final path integration yields the generating functional for Green functions,
\begin{equation}
  \label{eq:fpi}
  {\mathcal Z}(j,J)=\exp{\left[i\int d^2x\left(J_\Phi\hat{\Phi}(j)+J^1\hat{Y_1}(j)+J^2\hat{Y_2}(j)+L_{\rm amb}(j)\right)\right]}\,,
\end{equation}
where $J^1,J^2$ are appropriate linear combinations of $J,\bar{J}$
(cf.~\eqref{eq:redef}) and $\hat{\Phi}$, $\hat{Y}_1$ and $\hat{Y}_2$ are the
solutions obtained from \eqref{eq:sol2}-\eqref{eq:sol3} in the presence of
sources and for a certain choice of the integration constants, in particular
${\mathcal C}_0$. Ambiguities which result from a careful definition of the
first three terms in \eqref{eq:fpi} are collected in $L_{\rm amb}(j)$. Their
explicit form is not relevant for the current work. The curious reader may
wish to consult either section 7 of \cite{Grumiller:2002nm} or some of the
original literature \cite{Kummer:1998zs,Grumiller:2000ah} where these terms are discussed in detail. We just recall a heuristic argument showing
their necessity: it is seen clearly from \eqref{eq:fpi} that their absence
would imply 
\begin{equation}
  \label{eq:dasistschonwiedereinegleichungmiturlangemlabelhurra}
  \langle e^a\rangle=\left.\frac{\delta \ln{\mathcal Z}}{\delta
  j_a}\right|_{j=J=0}=0=\left.\frac{\delta \ln{\mathcal Z}}{\delta
  j_\omega}\right|_{j=J=0}=\langle\omega\rangle\,,
\end{equation}
because differentiation with respect to the sources $j$ and setting $J=j=0$
yields no contribution from the first three terms of the r.h.s.~of
\eqref{eq:fpi}. Thus, $L_{\rm amb}$ encodes the expectation values of
connection and Zweibeine.

A final remark is in order: because the theory does
not allow propagating physical modes the generating functional for Green
functions \eqref{eq:fpi}  solely\footnote{Of course, as discussed before it
  also depends on various integration constants, most of which may be absorbed
  by trivial redefinitions -- so essentially it depends on ${\mathcal C}_0$
  besides the sources. However, this dependence is a parametric and not a
  functional one.} depends on the external sources, and it does so in a very
specific way: the sources $J$ appear only linearly, while the dependence on
$j$ generically is non-polynomial. This has important consequences for
clustering properties of correlators. For instance, expectation values of the
form $\langle f(\Phi,Y_1,Y_2)\rangle$, where $f$ is an arbitrary function of
$\Phi,Y_1$ and $Y_2$, are just given by their classical value
$f(\Phi,Y_1,Y_2)$ (``the correlators cluster''), while expectation values
containing at least one insertion of either Zweibein or connection generically
receive non-trivial corrections in addition to the classical terms. This point
will be considered in more detail below. A particular consequence for the
Casimir \eqref{C} is 
\begin{equation}
  \label{eq:CC0}
  \langle{\mathcal C}\rangle={\mathcal C}_0\,,
\end{equation}
where $\mathcal{C}_0$ is the integration constant appearing in
(\ref{eq:sol3}). By choosing a particular value of ${\mathcal C}_0$ one fixes
the vacuum state.

\subsection{Solution in absence of sources}
In the absence of sources the solutions 
of (\ref{eq:eom1}) - (\ref{eq:eom3}) simplify to
\begin{align}
& Y_1= Y_1^h\,, \label{eq:sol2a}\\
& Y_2=\pm\sqrt{4|c|^2e^{-Q(\Phi)}\left({\mathcal C}-w(\Phi)\right)
-(Y_1^h)^2}\,, \label{eq:sol3a} \\
& \int^\Phi \frac{d\phi}{Y_2(\phi)} = x^1-\bar{x}^1\,,\label{eq:sol1a}
\end{align}
where $\bar{x}_1\in\mathbb{R}$ is the third integration constant. It can
always be absorbed by a trivial shift of the coordinate $x^1$ and thus may be
set to zero. For models without torsion ($Q=\rm const.$) or for Euclidean Ground
state models (defined by $e^{-Q(\Phi)}w(\Phi)=\rm const.$) the integration
constant $Y_1^h$ may be absorbed into a redefinition of $c$ (apart from
certain singular values). In these cases only one essential integration
constant remains, namely the Casimir $\mathcal C$. As a technical sidenote,
depending on the specific choice of the potentials $U,V$ (which then determine
$w,Q$) the integration of \eqref{eq:sol1a} in terms of known functions may be
possible. The Liouville model for $\alpha=0$ will be treated below as a particular example.

The existence of two branches of the square root in \eqref{eq:sol3a}
represents a delicate point. The crucial observation here is that, apart from diffeomorphisms, our gauge group actually is $O(2)$ which differs from $SO(2)$ by a discrete symmetry, namely reflections. Another way to see this fact is that the gauge orbits $Y_1^2+Y_2^2=\rm const.$ are circles for Euclidean signature (as opposed to hyperbolas for Minkowskian signature);\footnote{That is the reason why this issue never arose in the Minkowskian case. In Minkowskian space the solutions of $Y^+Y^-=\rm const.$ are hyperbolas and thus two branches exist. Fixing e.g.~$Y^+=+1$ determines not only $Y^+$ uniquely, but also the sign of $Y^-$! In contrast, for Euclidean signature the solutions $Y_1^2+Y_2^2=\rm const.$ are circles and fixing $Y_1=+1$ determines only $Y_1$ uniquely without restricting the sign of $Y_2$.} thus, fixing the value of $Y_1$ by choosing some $Y_1^h$ and fixing $|Y_2|$ by (\ref{eq:sol3}) is not enough---one encounters two ``Gribov copies''. Suppose for simplicity $c\in\mathbb{R}$. Then, even after fixing the gauge according to \eqref{eq:efeuk} one may still perform a residual discrete gauge transformation, namely exchanging all barred with unbarred quantities, corresponding to complex conjugation. Because the action is real it is invariant, and the gauge \eqref{eq:efeuk} for real $c$ is also invariant.\footnote{ If $c$ is not real then the argument becomes technically more complicated but in essence it remains the same, i.e., one still has to fix the reflection ambiguity.} But $Y_2\to-Y_2$ under this discrete transformation. Thus, fixing the sign in \eqref{eq:sol3a} removes this residual gauge freedom.
 
Nevertheless, in practice the best strategy is to keep both signs, to evaluate $\Phi$ for both and to choose whether one would like to have $\Phi$ monotonically increasing ($Y_2>0$) or decreasing ($Y_2<0$) as a function of $x^1$ at a certain point---e.g.~in the asymptotic region, whenever this notion makes sense.

\subsection{Example: Liouville model with $\alpha=0$}

For the Liouville model (\ref{UVLi}) with $\alpha=0$ there exists an
alternative way to solve the system \eqref{eq:eom1}-\eqref{eq:eom3}. As (\ref{eq:eom3}) does not depend on $\Phi$ one can simply solve this equation without invoking the conservation law and plug the result directly into (\ref{eq:eom1}). Of course, both methods yield the same result:
\begin{align}
& Y_1=Y_1^h\,,\label{eq:y1liou}\\
\label{eq:x2liou}
& Y_2 = - \frac{1}{d}\tan{(fx^1)}\,,\\
\label{eq:philiou}
& \Phi = \frac2a \ln{\left(\cos{(fx^1)}\right)}\,.
\end{align}
with
\begin{equation}
d:= \frac{a}{f}\,,\quad f:=\sqrt{a(a(Y_1^h)^2+4|c|^2b)}\,.
\end{equation}
It has been assumed that $a,b>0$ and thus $\mathcal C$ has to be positive. Integration constants have been absorbed by shifts of $x^1$ and $\phi$. The sign of $Y_2$ has been chosen such that $\Phi\to-\infty$ in the ``asymptotic region'' $fx^1=\pi/2$. In accordance with the previous discussion the asymptotic point $\Phi=-\infty$ may be added which yields the topology of a sphere. Note that for negative $a$ the quantity $Y_2$ has to be positive; this change of sign is in accordance with the first remark in \ref{sec:laulimamelina} because a change of the sign of $\Phi$ also changes the sign of $Y_2$ according to \eqref{eq:eom1}.

\subsection{Local quantum triviality versus correlators}
Local quantum triviality can be derived by looking at the effective action in
terms of mean fields. It can be checked easily that the effective action, up
to boundary contributions, is equivalent to the classical action in the gauge
(\ref{eq:efeuk}). We denote
\begin{equation}
  \label{eq:thislablisyourlabelthislabelismylabelfromlefttorightboundaryofthemonitor}
  \langle\omega_2\rangle=\frac{\delta \ln{{\mathcal Z}(j,J)}}{\delta j_\omega}
\end{equation}
by $\omega_2$, where ${\mathcal Z}$ is given by \eqref{eq:fpi}, and similarly for all other variables. One obtains by virtue of (\ref{eq:eom1})-(\ref{eq:eom3})
upon substitution of the sources  into$\int d^2x(j_\omega \omega_2+j_ie^i_2)$ (the other terms cancel trivially or
 yield boundary contributions)\footnote{The explicit form of $L_{\rm amb}$ mentioned above is neither relevant for the discussion of local
 quantum triviality nor for the special correlators considered below and thus we do not include it here.} 
 the effective Lagrangian density
\begin{equation}
  \label{eq:effact}
 L_{\rm eff.}= -(\partial_1\Phi -Y_2)\omega_2-\partial_1Y_1 e^1_2-(\partial_1Y_2+(Y_1^2+Y_2^2)U(\Phi)+2|c|^2 V(\Phi))e^2_2\,.
\end{equation}
Recalling the redefinitions (\ref{eq:redef}) and corresponding ones for the
Zweibeine ($e^1_2:=(ce_2^\ast+c^\ast e_2)/(2|c|^2)$,
$e_2^2:=i(ce_2^\ast-c^\ast e_2)/(2|c|^2)=-(\det e)/(2|c|^2)$) it is seen that
(\ref{eq:effact}) is nothing else than the classical action in the gauge (\ref{eq:efeuk}); to this end it is helpful to rewrite (\ref{eq:effact}) (up to boundary terms) as
\begin{multline}
  \label{eq:effact2}
L_{\rm eff.}=  Y_1 \partial_1e^1_2+Y_2(\partial_1 e^2_2 + \omega_2)+\Phi\partial_1\omega_2\\
+\det e((Y_1^2+Y_2^2)U(\Phi)/(2|c|^2)+V(\Phi))\,.
\end{multline}
With (\ref{eq:redef}) this reproduces exactly (\ref{eq:gfa}). Therefore, as
expected from the Min\-kow\-skian case the theory is locally quantum trivial.

Local quantum triviality by no means implies
that all correlators are trivial \cite{Bergamin:2004us}. For instance\footnote{Expectation values involving the $e_a$ and
  $\omega$ alone would require the inclusion of $L_{\rm amb}$.}, the Lorentz invariant 1-forms $Y^a(x)e_a(x)$ and $\epsilon_{ab}Y^a(x)e^b(x)$ exhibit interesting behaviour if delocalised
\begin{align}
\label{eq:cor1}
& \langle Y^a(x)e_a(y) \rangle = \langle Y^a(x)\rangle\langle e_a(y)\rangle +
\delta_1(x,y)\,,\\
\label{eq:cor2}
& \langle\epsilon_{ab}Y^a(x)e^b(y)\rangle = \epsilon_{ab}\langle Y^a(x)\rangle\langle e^b(y)\rangle + \delta_2(x,y)\,,
\end{align}
i.e.~the quantities $\delta_1$, $\delta_2$ will be shown to be nonvanishing,
while the remaining terms just yield the classical result expected from local
quantum triviality discussed above. It is recalled that the state in
\eqref{eq:cor1} and \eqref{eq:cor2} is fixed by choosing the integration
constant $\mathcal{C}_0$.
One may wonder why we take specifically the correlators \eqref{eq:cor1},
\eqref{eq:cor2}. The reason for this is twofold: first, they are Lorentz
invariant locally in the coincidence limit and still retain global Lorentz
invariance even non-locally.\footnote{By insertion of a Wilson line
  $\exp{i\int_x^y \omega^a{}_b}$ integrated over a certain path starting at
  the point $x$ and ending at the point $y$, one can construct correlators
  similar to the ones in \eqref{eq:cor1}, \eqref{eq:cor2} which are then
  locally Lorentz invariant. However, they do not allow an immediate physical
  interpretation as known from other topological models. Since the main message we want to convey is that the quantities
  $\delta_{1,2}$ are non-vanishing we only present the simpler calculation of
  the correlators \eqref{eq:cor1}, \eqref{eq:cor2}.} Second, correlators which
contain arbitrary powers of $\Phi,Y_1,Y_2$ alone cluster decompose into products of their classical values, so one needs at least one gauge field insertion. So the correlators above are the simplest non-trivial examples of non-local correlators which retain at least global Lorentz invariance.
Finally, it should be mentioned that the objects 
\begin{align}
& \delta_1(x,y)=\frac{\delta}{\delta j_1(y)} Y_1(x)+\frac{\delta}{\delta j_2(y)} Y_2(x)\,, \label{eq:d1} \\
& \delta_2(x,y)=\frac{\delta}{\delta j_2(y)} Y_1(x)-\frac{\delta}{\delta j_1(y)} Y_2(x) \label{eq:d2}
\end{align} 
are 1-forms and thus naturally may be integrated along a path.

To determine $\delta_1$ and $\delta_2$ one needs the variations $\delta Y_i(x)/ \delta j_j(y)$
with $i,j=1,2$ and $Y$ being the solution \eqref{eq:sol2} and \eqref{eq:sol3},
resp. The state is defined so that (\ref{eq:CC0}) holds, thus fixing the
integration constant in (\ref{eq:sol3}).
For $i=1$ the expressions follow straightforwardly from \eqref{eq:sol2}:
\begin{align}
  \label{eq:lu2}
  \frac{\delta}{\delta j_1(y)} Y_1(x) &= \int_{\bar{x}^1}^{x^1} dz^1 \delta^2(y-z) &
  \frac{\delta}{\delta j_2(y)} Y_1(x) &= 0
\end{align}
The lower integration limit should be chosen  conveniently. To fix the remaining free parameters induced in this way
it is natural to require a vanishing result in the coincidence limit $x=y$. 
Following
the choice of \cite{Bergamin:2004us} we define $\theta(0)=1/2$ and choose $\bar{x}^1=y^1$. More involved is the
variation of $Y_2$ as---beside the explicit appearance of $j_i$---the dilaton implicitly
depends on the sources. Therefore the variation yields an integral
equation, in particular for $j_2$
\begin{equation}
\label{eq:lu3}
\begin{split}
  \left. \frac{\delta}{\delta j_2(y)} Y_2(x)\right|_{j=0} &= \frac{1}{2 Y_2}\Biggl(2 e^{-Q}
  \partial_1^{-1}{}_{xz}\bigl(e^Q \partial_1 \Phi \delta^2(y-z) \bigr) \\
  &\quad +
  \left. \left(\frac{\partial}{\partial \Phi} Y_2^2\right) \int^{x^1} dz \frac{\delta}{\delta j_2(y)}
  Y_2(z) \Biggr)\right|_{j=0}\,.
\end{split}
\end{equation}
Its solution can be obtained straightforwardly as
\begin{multline}
  \label{eq:lu4}
    \frac{\delta}{\delta j_2(y)} Y_2(x) =  [e^Q Y_2]_y \biggl(
    \bigl[\frac{e^{-Q}}{Y_2}\bigr]_x + (\partial_{x^1} Y_2)
    \bigl(F(x)-F(y)\bigr) \biggr)\\ \bigl(\theta(x^1-y^1)-\frac12 \bigr) \delta(x^2-y^2)\,,
\end{multline}
where $F(x)$ is the abbreviation for
\begin{equation}
  \label{eq:lu5}
  F(x) = \int^{x^1} dz^1 \frac{e^{-Q}}{Y_2^2}\ .
\end{equation}

For sake of completeness it should be noted that a similar calculation can be done for the variation with respect to
$j_1$. This correlator depends on an additional integral
\begin{equation}
  \label{eq:lu7}
  G(x) = \int^{x^1}dz^1 Y_2^{-2}\ .
\end{equation}
Using the same prescription as the one that led to \eqref{eq:lu4} the result becomes
\begin{multline}
  \label{eq:lu8}
    \frac{\delta}{\delta j_1(y)} Y_2(x) =  Y_1^h \biggl\{\frac1{Y_2(x)}
    \bigl( e^{Q(y)-Q(x)} - 1\bigr)  + (\partial_{x^1}Y_2) \\
    \Bigl(e^{Q(y)}\bigl(F(x)-F(y)\bigr)- G(x) + G(y)\Bigr) \biggr\}
    \bigl(\theta(x^1-y^1)-\frac12 \bigr) \delta(x^2-y^2)\ .
\end{multline}

By a dilaton-dependent conformal transformation $e_\mu^a\to \tilde e_\mu^a=
e^{\rho(X)} e_\mu^a$ one can map one dilaton gravity model upon another.
The potentials $\tilde U$ and $\tilde V$ will be, of course, different
from the original ones $U$ and $V$ (cf.~e.g.~Eq.~(3.41) 
in ref.~\cite{Grumiller:2002nm}). This transformation
is usually called the transition to another conformal frame, although
models corresponding to different frames in general are already
inequivalent at the
classical level (because of possible singularities in the 
transformation).
By a suitable choice of $\rho (X)$
one can achieve $\tilde U=0$ (and, consequently, $\tilde Q=0$). 
Note that in a conformal frame
with $Q=0$ Eq.~\eqref{eq:lu8} vanishes identically. Thus, the quantity
$\delta_2(x,y)$ captures the dependence on the conformal frame and vanishes if
there is no kinetic term for the dilaton. In contrast, \eqref{eq:lu4} is
non-vanishing even for this simple case and \eqref{eq:lu2} yields a frame
independent contribution. Thus, $\delta_1(x,y)$ encodes both, frame
independent as discussed in \cite{Bergamin:2004us} as well as frame dependent information.\footnote{At this point we should clarify a misleading statement in ref.~\cite{Bergamin:2004us}: while everything is correct until Eq.~(4.37), Eq.~(4.38) and subsequent Eqs.~depending on it are only true for $Q=0$. In particular, the correlator Eq.~(4.42) in addition to the purely topological information discussed in that paper encodes also information about the conformal frame.}

\section{Conclusions}\label{se:conclu}
The methods which have been very successful in two-dimensional gravity \cite{Grumiller:2002nm} 
are extended to Euclidean signature models. To establish classical and
quantum integrability of dilaton gravities also in Euclidean space, is a
necessary prerequisite to treat strings and, in particular, Liouville
theory. We are able to explicitly construct all local classical solutions
for all theories of this type.
In quantum theory we perform the path integral nonperturbatively, where,
however, problems specific for Euclidean signature have to be
overcome: the
Eddington-Finkelstein gauge which plays a key role in the comparatively
simpler case of Minkowski space theories has no counterpart here, so that
a new type of complex gauge fixing had to be introduced. This again
permitted us to follow in broad lines the strategy, well-tested in the
Minkowski space. As in that case we observe \emph{local} quantum triviality.
Despite this property, non-local correlator functions exist.
In examples we show how to compute them explicitly in our approach. We,
therefore, believe to have performed a very important first step towards
the study of general correlators for the Liouville model which may be
related to matrix models and strings. This will be the subject of a future
publication.

\section*{Acknowledgements}
We thank H.\ Balasin for valuable comments. One of the authors (DVV) is grateful to S.~Alexandrov and V.~Kazakov for
fruitful discussions. DG and LB acknowledge the hospitality at the Vienna
University of Technology and at the University of Leipzig, respectively,
during the preparation of this work.

LB is supported by project P-16030-N08 of the Austrian Science Foundation
(FWF), DG by an Erwin-Schr\"odinger fellowship, project
J-2330-N08 of the Austrian Science Foundation (FWF) and in part by project
P-16030-N08 of the FWF. Part of the support of DVV is due to the DFG Project BO
1112/12-1 and to the Multilateral Research Project ``Quantum gravity,
cosmology and categorification'' of the Austrian Academy of Sciences and the
National Academy of Sciences of the Ukraine.

\appendix
\section{Equivalence of first and second order formulations}\label{AppA}
The proof of classical equivalence between first and second order formulations
differs in details only from the corresponding one for
Minkowski signature \cite{Katanaev:1997ni,Grumiller:2002nm}.
Quantum equivalence has been demonstrated in \cite{Kummer:1997hy}.

The action (\ref{act1}) is rewritten in the component form,
\begin{equation}
L= \frac 12 \int_{\mathcal{M}} d^2\sigma 
\left[ Y^a (D_\mu e^a_\nu )\tilde \varepsilon^{\mu\nu}
+\Phi \partial_\mu \omega_\nu \tilde \varepsilon^{\mu\nu}
+\dete \mathcal{V}(Y^2,\Phi)
\right]\,,\label{compact}
\end{equation}
where
\begin{equation}
\dete = \det e_\mu^a = -\frac 12 \varepsilon^{ab}
\tilde\varepsilon^{\mu\nu}e_\mu^a e_\nu^b \,. \label{dete}
\end{equation}

The two Levi-Civita symbols, $\varepsilon^{ab}$ and 
$\tilde\varepsilon^{\mu\nu}$, are defined as 
$\varepsilon^{12}=\tilde\varepsilon^{12}=1$. The one with
anholonomic indices ($\varepsilon^{ab}$) is tensorial, while
the symbol with holonomic indices ($\tilde\varepsilon^{\mu\nu}$) is a
tensor density.

The connection $\omega_\mu$ is split into the Levi-Civita part $\hat\omega$
and the torsion part $t_\mu$, $\omega_\mu=\hat\omega_\mu +t_\mu$.
By definition, the Levi-Civita connection
\begin{equation}
\hat\omega_\mu = \dete^{-1} e^a_\mu \tilde \varepsilon^{\rho\sigma}
(\partial_\rho e^a_\sigma ) \label{LCcon}
\end{equation}
corresponds to vanishing torsion:
\begin{equation}
\tilde\varepsilon^{\mu\nu} \left( \partial_\mu e^a_\nu +
\varepsilon^{ab} \hat\omega_\mu e^b_\nu \right) =0 \label{compatib}
\end{equation}

For $\mathcal{V} (Y^2,\Phi)$ given by (\ref{restrict}) the equation
of motion for $Y^a$ reads
\begin{equation}
t_\mu \tilde\varepsilon^{\mu\rho} +\dete U(\Phi) Y^a \varepsilon^{ac}
e^\rho_c =0 \,,\label{eomY}
\end{equation}
where $e^\rho_a$ are the inverse components of the ones in the Zweibein 1-form. This equation is linear and,
therefore, it may be substituted back into the action to eliminate $t$:
\begin{equation}
L= \frac 12 \int_{\mathcal{M}} d^2\sigma 
\left[ \Phi \partial_\mu \hat \omega_\nu \tilde \varepsilon^{\mu\nu}
+\dete U(\Phi ) \left( (\partial_\rho \Phi )e_c^\rho \varepsilon^{ac} Y^a
-\frac 12 Y^2 \right) +\dete V(\Phi ) \right] \,.\label{comac2}
\end{equation}

The equation of motion for $Y^a$ following from this action is again
linear:
\begin{equation}
Y^a=(\partial_\rho \Phi) e_c^\rho \varepsilon^{ac}\,. \label{moreY}
\end{equation} 
We also note that 
\begin{equation}
\partial_\mu \hat \omega_\nu \tilde\varepsilon^{\mu\nu} = -\dete \frac R2\,,
\label{homR}
\end{equation}
where $R$ is the curvature scalar. With our sign conventions $R=2$
on unit $S^2$. Next we substitute (\ref{moreY}) and (\ref{homR}) in
(\ref{comac2}). We also note that we can re-express everything in terms
of the Riemannian metric $g_{\mu\nu}= e_\mu^ae^b_\nu \delta_{ab}$ 
instead of $e_\mu^a$.
As a result, we obtain the second order action (\ref{dilact}) (up to
an irrelevant overall factor of $1/2$).

\section{Compact manifolds}\label{AppB}
On a compact connected
manifold $\mathcal{M}$ the Laplace operator $\Delta$
has a zero mode. Consequently, the conformal anomaly (\ref{Tmm}) has to be
modified, as well as the local and non-local actions (\ref{locP}) and 
(\ref{nonlocP}). This procedure was considered in detail by Dowker
\cite{Dowker:1994rt}. One has to subtract the zero mode
contribution from (\ref{Tmm}) so that the modified expression
for the anomaly reads:
\begin{equation}
T_\mu^\mu = \frac {N-1}{24\pi} [ R + \tilde \mu ] 
-\frac{N-1}{\mathcal{A}} \,,\label{cTmm}
\end{equation}
where $\mathcal{A}$ is the area of $\mathcal{M}$.

The non-local action (\ref{nonlocP}) receives several additional terms
\cite{Dowker:1994rt}, and the Liouville action (\ref{locP}) should contain a
non-local term $\frac{N-1}2 \ln \mathcal{A}$. Accordingly, one has to
add to the first-order action (\ref{act1}) the term:
\begin{equation}
L_{\mathrm{area}}=\beta \ln \mathcal{A} \,,\label{areact}
\end{equation}
where the constant $\beta$ can be fixed by the considerations presented
above, but we consider here general values of $\beta$. A more convenient
representation uses an auxiliary variable $h$:
\begin{equation}
\tilde L_{\mathrm{area}}=\beta (h\mathcal{A} -1 - \ln h ) \,.\label{hact}
\end{equation}

The action (\ref{locP}) is the only one directly derived
from strings. It corresponds to $\alpha =0$ in the potentials (\ref{UVLi}).
At the classical level, for that value of $\alpha$ adding the term (\ref{hact}) results in
shifting $b$ to
\begin{equation}
\hat b = b+\beta h \,.\label{hatb}
\end{equation}
Therefore, all solutions obtained in sec.\ \ref{sLi} remain valid
after the replacement $b\to \hat b$. The exact value of this shift is
fixed by a ``global'' equation of motion following from
(\ref{hact}):
\begin{equation}
\mathcal{A}=1/h \,.\label{A1h}
\end{equation}
For the present model (cf. Case sI of sec.\ \ref{sLi}), we have
\begin{equation}
\mathcal{A}=\theta_0 \int_{-\infty}^{\Phi_0} e^{a\Phi} d\Phi =
\frac{2\pi a}{\hat b} \,.\label{AVI}
\end{equation}
We recall that for infinite volume solutions no modification of the
action is needed. Equations (\ref{A1h}) and (\ref{AVI}) yield
\begin{equation}
h=\frac{b}{2\pi a -\beta} \,.\label{hab}
\end{equation}

Another problem arising for compact manifolds is that the equation of motion
for $\Phi$ from the action \eqref{locP},
\begin{equation}
R=-2\Delta\Phi\,, \label{RDP}
\end{equation}
has no smooth solution except for the case of vanishing Euler characteristic.
In our approach this problem is resolved automatically, because the classical
solution obtained in Case sI becomes compact after adding a point
with $\Phi =-\infty$.
This solution is even unique as we identify $\Phi$ with a coordinate
on $\mathcal{M}$. Of course, a similar problem exists also in quantum
theory where it should be solved by choosing appropriate boundary conditions.





\providecommand{\href}[2]{#2}\begingroup\raggedright\endgroup

\end{document}